\documentclass[12pt]{article}
\usepackage{epsfig}
\usepackage{amsmath}
\usepackage{hhline}
\usepackage{amssymb}
\usepackage{times}
\usepackage{color}
\usepackage{lineno}

\newlength{\dinwidth}
\newlength{\dinmargin}
\newcommand{\QQ}  {\mbox{${Q^2}$}}

\begin{document}

\pagestyle{empty}
\begin{titlepage}

\noindent
\begin{flushleft}
DESY 05-249\hfill ISSN 0418-9833\\
December 2005
\end{flushleft}

\vspace{2cm}

\begin{center}
\begin{Large}
{\bf  First Measurement of Charged Current Cross Sections at HERA
      with Longitudinally Polarised Positrons}

\vspace{2cm}

H1 Collaboration

\end{Large}
\end{center}

\vspace{2cm}

\begin{abstract}

\noindent
Data taken with positrons of different longitudinal polarisation states
in collision with unpolarised protons at HERA are used to measure the
total cross sections of the charged current process, $e^+p\rightarrow
\overline{\nu}X$, for negative four-momentum transfer squared
$Q^2>400\,{\rm GeV}^2$ and inelasticity $y<0.9$.  
Together with the corresponding cross
section obtained from the previously published unpolarised
data, the polarisation dependence of the charged current cross
section is measured for the first time at high $Q^2$ and 
found to be in agreement with the Standard Model prediction.  

\end{abstract}

\vspace{1.0cm}

\begin{center}
Dedicated to the memory of our dear friend and colleague, Alexei~Babaev \\

\vspace{1cm}

Submitted to Phys. Lett. {\bf B}

\end{center}

\end{titlepage}


A.~Aktas$^{9}$,                
V.~Andreev$^{25}$,             
T.~Anthonis$^{3}$,             
B.~Antunovic$^{26}$,           
S.~Aplin$^{9}$,                
A.~Asmone$^{33}$,              
A.~Astvatsatourov$^{3}$,       
A.~Babaev$^{24,\dag}$,              
S.~Backovic$^{30}$,            
J.~B\"ahr$^{38}$
% ...

\bigskip{\it
% ...
$ ^{3}$ Inter-University Institute for High Energies ULB-VUB, Brussels;
          Universiteit Antwerpen, Antwerpen; Belgium$^{ c}$ \\
% ...
$ ^{9}$ DESY, Hamburg, Germany \\
% ...
$ ^{24}$ Institute for Theoretical and Experimental Physics,
          Moscow, Russia$^{ l}$ \\
 $ ^{25}$ Lebedev Physical Institute, Moscow, Russia$^{ e}$ \\
 $ ^{26}$ Max-Planck-Institut f\"ur Physik, M\"unchen, Germany \\
% ...
$ ^{30}$ Faculty of Science, University of Montenegro,
          Podgorica, Serbia and Montenegro$^{ e}$ \\
% ...
$ ^{33}$ Dipartimento di Fisica Universit\`a di Roma Tre
          and INFN Roma~3, Roma, Italy \\
% ...
$ ^{38}$ DESY, Zeuthen, Germany \\
% ...
$ ^{\dag}$ Deceased\\

\bigskip
% ...
$ ^c$ Supported by FNRS-FWO-Vlaanderen, IISN-IIKW and IWT
      and  by Interuniversity
Attraction Poles Programme,
      Belgian Science Policy \\
% ...
$ ^e$ Supported by the Deutsche Forschungsgemeinschaft \\
% ...
$ ^l$ Partially Supported by Russian Foundation
      for Basic Research,  grants  03-02-17291
      and  04-02-16445 \\
}

\newpage

\pagestyle{plain}

\section{Introduction}
In autumn 2003 the HERA accelerator started operation of the second
phase of its $ep$ collider programme. The $e^+p$ data collected by the
H1 and ZEUS experiments since then were taken with a longitudinally
polarised positron beam for the first time. Measurements of deep
inelastic scattering (DIS) with polarised leptons on protons allow the
parton distribution functions (PDFs) of the proton to be further
constrained through polarisation asymmetries~\cite{klein} and specific
tests of the electroweak (EW) parts of the Standard Model to be
performed~\cite{ewpaper,ewheraproc}.  In particular, the measurements
presented here extend the tests of the ${\rm V}-{\rm A}$ structure of
charged current interactions from low $Q^2$~\cite{charm} into the high
$Q^2$ regime, where $Q^2$ is the negative four-momentum transfer
squared.

At HERA DIS proceeding via charged currents (CC), \mbox{$ep\rightarrow \nu X$},
and neutral currents (NC), \mbox{$ep \rightarrow eX$}, can be measured 
accurately~\cite{h1hiq2,h1lowq2}. 
The polarisation dependence of the CC and NC cross sections is
fixed within the Standard Model framework. Specifically, the Standard
Model predicts, from the absence of right handed charged currents, that the
CC $e^+p$ cross section is directly proportional to the fraction of right
handed positrons in the beam.

In this paper first measurements of the charged current total
cross sections, $\sigma^{\rm tot}_{\rm CC}$, are reported for two values of
longitudinal polarisation, $P_e=(N_R-N_L)/(N_R+N_L)$,
with $N_R$ ($N_L$) being the number of right (left) handed positrons
in the beam. The corresponding data sets are termed the $R$ and $L$
data sets respectively.  The $R$ data set has a luminosity weighted
mean polarisation value of $(33.6\pm0.7)\,\%$ and an integrated
luminosity value of $26.9\pm 0.6\,{\rm pb}^{-1}$.  The corresponding
numbers for the $L$ data set are $(-40.2\pm 1.1)\,\%$ and $20.7\pm
0.5\,{\rm pb}^{-1}$.  In both data sets the incident positron beam
energy is $27.5\,{\rm GeV}$, whilst the unpolarised proton beam energy
is $920\,{\rm GeV}$.  This yields a centre-of-mass energy of
$\sqrt{s}=318\,{\rm GeV}$.  

The measurements presented here, as well as the corresponding one obtained 
using the published unpolarised data, are compared to Standard Model 
expectations and a linear fit to $\sigma^{\rm tot}_{\rm CC}$ 
as a function of $P_e$ is performed. 
The result of the fit is used to derive a
cross section for a fully left handed positron beam corresponding to $P_e = -1$.

\section{Charged Current Cross Section}
The measured double differential CC cross section for collisions of polarised 
positrons with unpolarised protons, corrected for QED radiative
effects, may be expressed as
%
\begin{equation}
\frac{{\rm d}^2\sigma_{\rm CC}}{{\rm d}x {\rm d}\QQ} = (1+P_e)
 \frac{G_F^2}{4\pi x } \left[\frac{M_W^2}{M_W^2+Q^2} \right]^2
\left(Y_+W_2-Y_-xW_3-y^2W_L \right ) \cdot (1+\delta^{\rm CC}_{\rm weak})\,,
\label{Scc}
\end{equation}
where $x$ is the Bjorken $x$ variable and $y$ characterises the
inelasticity of the interaction. The Fermi constant $G_F$ is 
defined~\cite{hector} using the weak boson masses.
Other quantities in Eq.(\ref{Scc}) include $M_W$, 
the mass of the $W$ boson, $W_2$,
$xW_3$ and $W_L$, CC structure functions, and $\delta^{\rm CC}_{\rm weak}$,
the weak radiative corrections.  The helicity dependences of the weak
interaction are contained in $Y_\pm=1\pm(1-y)^2$. In the quark parton
model (QPM), where $W_L\equiv 0$, the structure functions $W_2$ and
$xW_3$ for $e^+p$ scattering may be expressed as the sum and
difference of the quark and anti-quark momentum distributions,
$xq(x,Q^2)$ and $x\overline{q}(x,Q^2)$:
\begin{eqnarray}
W_2  &=& x(\phantom{-}\overline{u}+\overline{c}+d+s)\,,\\
xW_3 &=& x(-\overline{u}-\overline{c}+d+s)\,.
\end{eqnarray}
The total cross section, $\sigma^{\rm tot}_{\rm CC}$, is defined as the
integrated cross section in the kinematic region $Q^2>400\,{\rm
GeV}^2$ and $y<0.9$.  From Eq.(\ref{Scc}) it can be seen that the
cross section has a linear dependence on the polarisation of the
positron beam $P_e$. For a fully left handed positron beam, $P_e=-1$, 
the cross section is identically zero in the Standard Model.

\section{Experimental Technique}
 
At HERA transverse polarisation of the positron beam arises naturally
through synchrotron radiation via the Sokolov-Ternov
effect~\cite{spin}. In $2000$ a pair of spin rotators was installed in
the beamline on either side of the H1 detector, allowing transversely
polarised positrons to be rotated into longitudinally polarised states and
back again. The degree of polarisation is constant around the HERA
ring and is continuously measured using two independent polarimeters
LPOL~\cite{lpol} and TPOL~\cite{tpol}.  The polarimeters are situated
in beamline sections in which the beam leptons have longitudinal and
transverse polarisations respectively.  Both measurements rely on an
asymmetry in the energy spectrum of left and right handed circularly
polarised photons undergoing Compton scattering with the positron
beam. The TPOL measurement uses in addition a spatial asymmetry. The
LPOL polarimeter measurements are used when available and TPOL
measurements otherwise.

The H1 detector components most relevant to this analysis are the
liquid argon (LAr) calorimeter, which measures the positions and
energies of charged and neutral particles over the polar\footnote{The
polar angle $\theta$ is defined with respect to the positive $z$
axis, the direction of the incident proton beam.} angular range
$4^\circ<\theta<154^\circ$, and the inner tracking detectors, which
measure the angles and momenta of charged particles over the range
$7^\circ<\theta<165^\circ$. A full description of the detector can be
found in~\cite{h1det}.

Simulated DIS events are used in order to determine acceptance corrections.
DIS processes are generated using the DJANGO~\cite{django} Monte Carlo (MC)
simulation program, which is based on LEPTO~\cite{lepto} for 
the hard interaction and HERACLES~\cite{heracles} for single photon
emission and virtual EW corrections.
LEPTO combines ${\cal O}(\alpha_s)$ matrix elements with higher order QCD
effects using the colour dipole model as implemented in ARIADNE~\cite{cdm}.
The JETSET program~\cite{jetset} is used to simulate 
the hadronisation process.  
In the event generation the DIS cross section is calculated using the 
H1 PDF $2000$~\cite{h1hiq2} parametrisation for the proton PDFs.

The dominant $ep$ background contribution arises from photoproduction
processes. These are simulated using the PYTHIA~\cite{pythia} MC with
leading order PDFs for the proton taken from CTEQ~\cite{cteq5}
and for the photon from GRV~\cite{ggrv}. Further backgrounds from NC
DIS, QED-Compton scattering, lepton pair production, prompt photon
production and heavy gauge boson ($W^{\pm},Z^0$) production are also
simulated; their final contribution to the analysis sample is small. Further
details are given in~\cite{h1hiq2}.

The detector response to events produced by the generation
programs is simulated in detail using a program based on
GEANT~\cite{GEANT}. These simulated events are then subjected to the
same reconstruction and analysis chain as the data.

The selection of CC interactions follows closely that
of the previously published analysis of unpolarised data from H1~\cite{h1hiq2}
and is briefly described below.  
The CC events are characterised as having large unbalanced
transverse momentum, $P_{T,h}$, attributed to the undetected neutrino.
The quantity $P_{T,h}$ is determined from
$P_{T,h} = \sqrt{(\sum_i{p_{x,i}})^2+(\sum_i{p_{y,i}})^2}$, where
the summation is performed
over all particles of the hadronic final state. 
The hadronic final state particles are reconstructed using a combination of 
tracks and calorimeter deposits in an energy flow algorithm that avoids double
counting~\cite{bpthesis}.  

The CC kinematic quantities are determined from the hadonic final
state~\cite{jb} using the relations
\begin{equation}
   y_{h} = \frac{E_h-p_{z,h}}{ 2 \ E_e }\,,
   \hspace*{2cm}
   Q^2_{h} = \frac{P_{T,h}^2}{ 1-y_{h}}\,,
   \hspace*{2cm}
    x_h=\frac{Q^2_h} {s \ y_h}\,,
\end{equation}
where $E_h-p_{z,h}\equiv \sum_i (E_i-p_{z,i})$ and $E_e$ is the incident positron
beam energy.

NC interactions are also studied as they provide an accurate and high
statistics data sample with which to check the detector response.  The
selection of NC interactions is based mainly on the requirement of an
identified scattered positron in the LAr calorimeter, with an energy
$E^\prime_e>11\,{\rm GeV}$.  The NC sample is used to carry out an
{\it in-situ} calibration of the electromagnetic and hadronic energy
scales of the LAr calorimeter using the method described
in~\cite{h1hiq2}. The hadronic calibration procedure is based on the
balance of the transverse energy of the positrons with that of the
hadronic final state.
The calibration procedure gives good agreement between data
and simulation within an estimated uncertainty of $2\%$.

In addition, NC events are used for studies of systematic
uncertainties in the charged current analysis.  The data are processed
such that all information from the scattered positron is suppressed,
providing the so-called {\it pseudo-CC}
sample~\cite{bpthesis,adilthesis,ringaile}. This sample mimics CC
interactions allowing trigger and selection efficiencies to be checked
with high statistical precision and independently of the MC simulation.

\section{Measurement Procedure}

Candidate CC interactions are selected by requiring $P_{T,h}>12\,{\rm
GeV}$ and a reconstructed vertex within $35\,{\rm cm}$ in $z$ of the
nominal interaction point. In order to ensure high efficiency of the
trigger and good kinematic resolution the analysis is further
restricted to the domain of 
$0.03<y_h<0.85$.  The $ep$ background is dominantly due to
photoproduction events, in which the scattered positron escapes
undetected in the backward direction and transverse momentum
is missing due to fluctuations in the detector response or
undetected particles. This background is suppressed exploiting the
correlation between $P_{T,h}$ and the ratio $V_{ap}/V_p$ of 
transverse energy flow
anti-parallel and parallel to the hadronic final state transverse
momentum vector~\cite{bpthesis,adilthesis,ringaile}.
The suppression cuts are different for the $R$ and $L$ data sets as the
relative photoproduction contributions differ in the two samples.  The
residual $ep$ background is negligible for most of the measured
kinematic domain.  The simulation is used to estimate this
contribution, which is subtracted statistically from the CC data
sample. Non-$ep$ background is rejected by searching for
typical cosmic ray and beam-induced background event
topologies~\cite{bpthesis,adilthesis,ringaile}. 
The final $R$ ($L$) CC data sample amounts
to $\simeq 700$ ($\simeq 200$) events.

\begin{figure}[htb]
\begin{center}
\begin{picture}(90,120)(0,0)
\setlength{\unitlength}{1 mm}
\put(-25,50){\epsfig{file=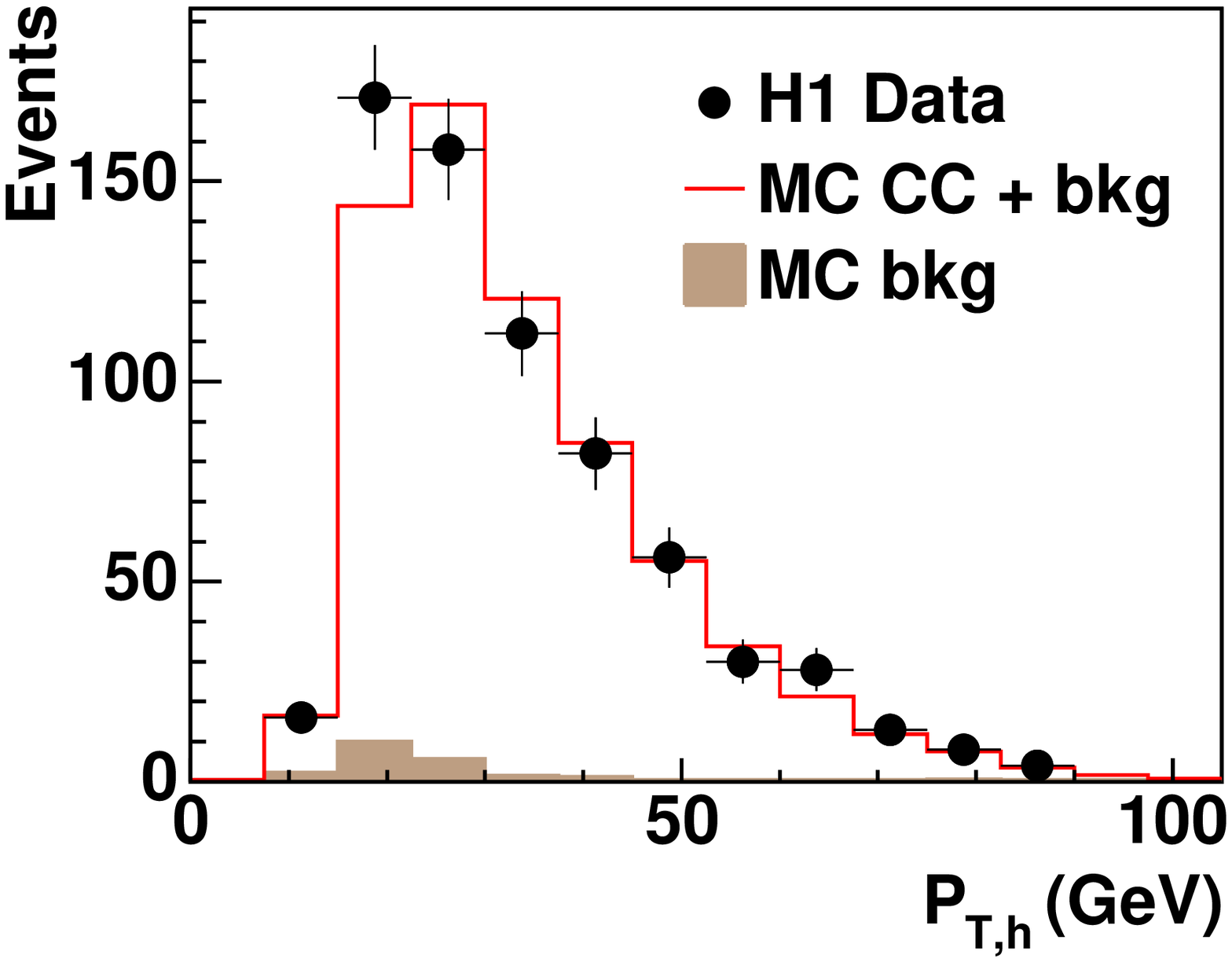,width=7.5cm}}
\put(50,50){\epsfig{file=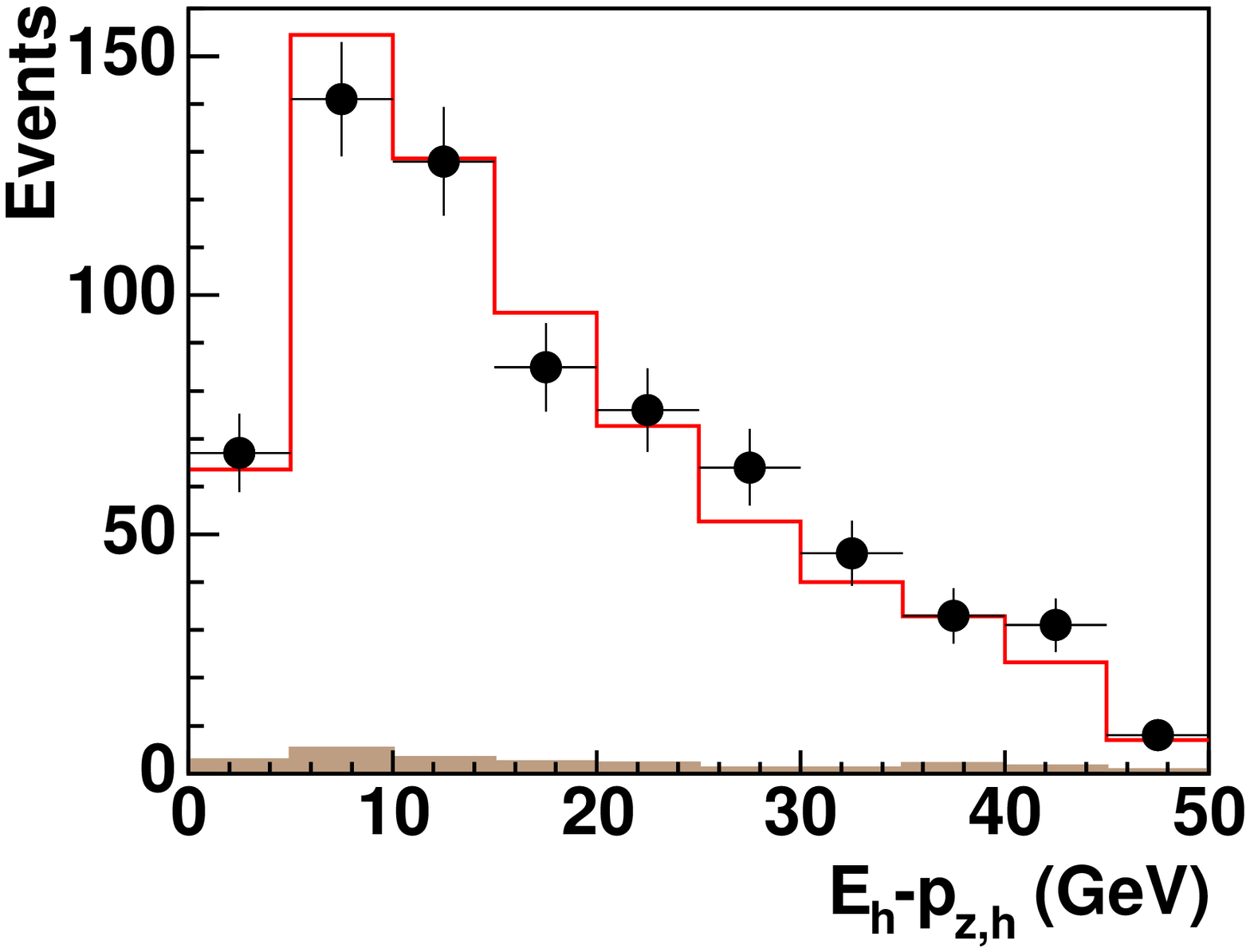,width=7.5cm}}
\put(-25,-5){\epsfig{file=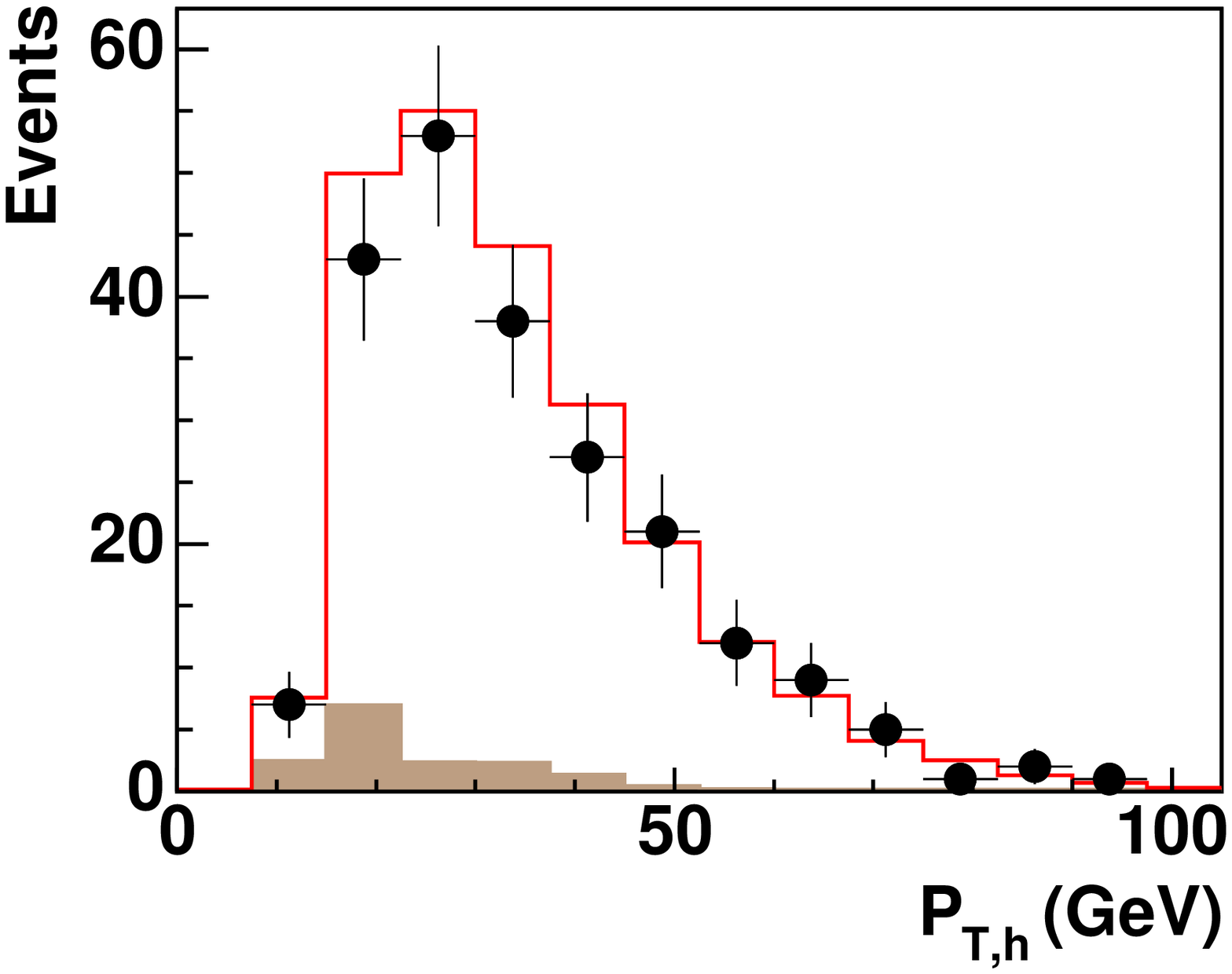,width=7.5cm}}
\put(50,-5){\epsfig{file=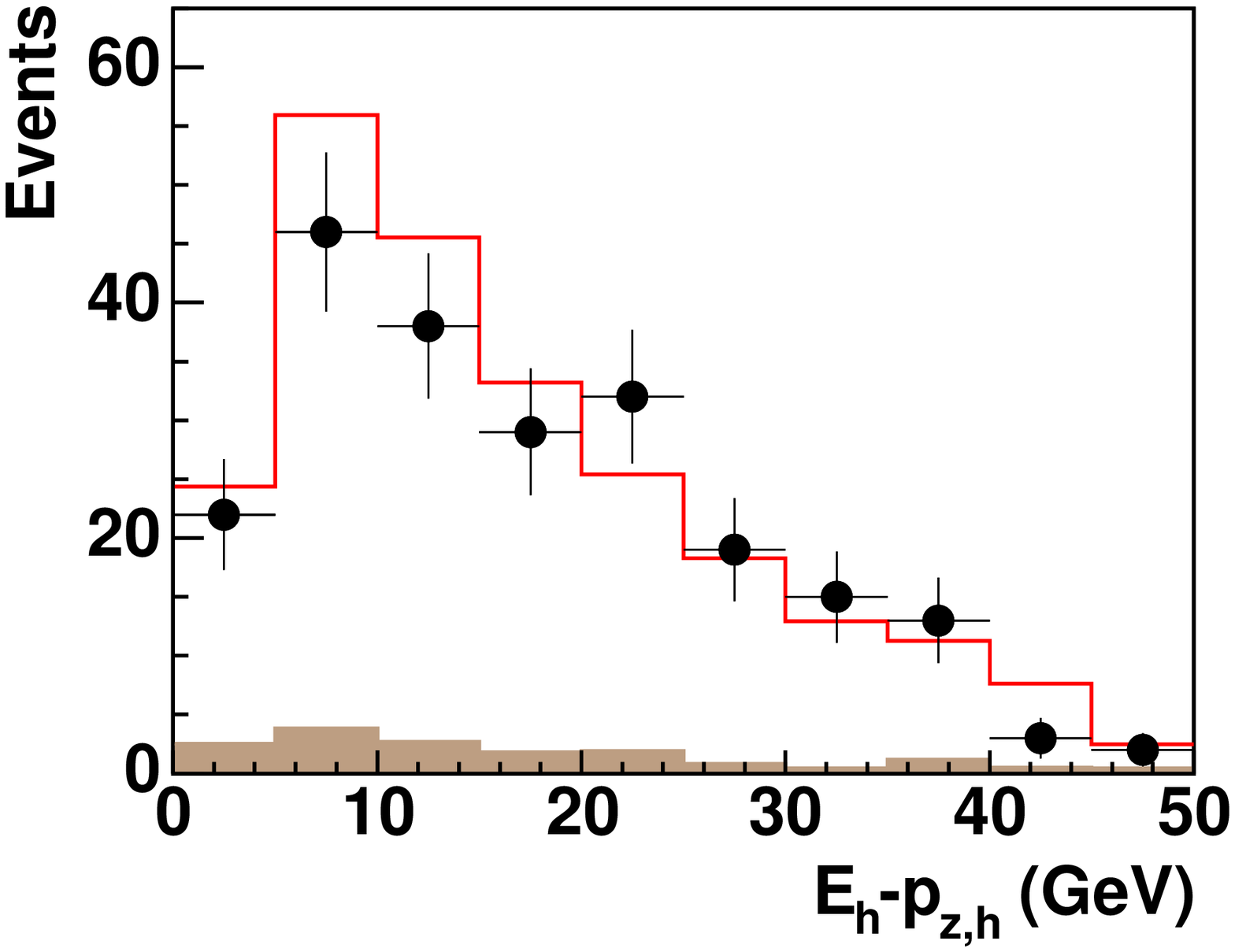,width=7.5cm}}
\put( 35,75){\bf (a)}
\put(110,75){\bf (b)}
\put( 35,25){\bf (c)}
\put(110,25){\bf (d)}

\put( 40,97){\bf R}
\put(115,97){\bf R}
\put( 40,40){\bf L}
\put(115,40){\bf L}

\end{picture}
\end{center}
\caption{
  Distributions of $P_{T,h}$ (a,c) and $E_h-p_{z,h}$ (b,d) for
  the selected events in the right handed (R) and left handed (L) 
  data sets. The Monte Carlo (MC)
  contributions from the charged current (CC) process and the
  $ep$ background (bkg) processes are shown as open histograms with
  the latter contribution alone being shown as shaded histograms.}
\label{cc}
\end{figure}

The $P_{T,h}$ and $E_h-p_{z,h}$ distributions of the selected events
are shown in Fig.\,\ref{cc}a,b for the $R$ sample and in
Fig.\,\ref{cc}c,d for the $L$ sample. The simulation provides a
good description of the data. The contribution of background
photoproduction processes is small and has the largest influence at
low $P_{T,h}$.

Events with $Q^2_h> 400\,{\rm GeV}^2$ are used to measure the cross sections,
which correspond to the kinematic region
$Q^2>400\,{\rm GeV}^2$ and $y<0.9$ and thus are corrected for the effects
of the analysis cuts.
The correction factor is calculated to be $1.067$ using the H1 PDF $2000$
parametrisation.

The systematic uncertainties on the cross section measurements are
discussed briefly below (see \cite{bpthesis,adilthesis,ringaile} and
references therein for more details). Positive and negative
variations of one standard deviation of each error source are found to
yield errors which are symmetric to a good approximation. The
systematic uncertainties of each source are taken to be fully
correlated between the cross section measurements unless stated
otherwise.

\begin{itemize}
\item An uncertainty of $2\%$ is assigned to the hadronic energy
  measured in the LAr calorimeter, of which $1\%$ is considered as a
  correlated component to the uncertainty. This results in a
  total uncertainty of $1.3\%$ on the cross section measurements.
    
\item A $10\%$ uncertainty is assigned to the amount of energy
  in the LAr calorimeter attributed to noise, which gives rise to a
  systematic error of $0.3\%$ on the cross section measurements.
  
\item The variation of cuts against photoproduction on 
  $V_{ap}/V_p$ and $P_{T,h}$ has an effect on the cross sections of $0.6\%$.
  
\item A $30\%$ uncertainty 
  on the subtracted $ep$ background is determined from a comparison of
  data and simulation after relaxing the anti-photoproduction cuts,
  such that the sample is dominated by photoproduction events. 
  This results in a systematic
  error of $0.5\%$ ($1\%$) on the cross section of the $R$ ($L$)
  data.
  
\item The non-$ep$ background finders introduce an inefficiency for 
  CC events. The associated uncertainty is estimated using pseudo-CC data and
  found to depend on $y$. An uncertainty of $2\%$ is applied for
  $y<0.1$ and $1\%$ for $y>0.1$. This yields an uncertainty of $1\%$
  on the cross section measurements.

\item A $y$-dependent error is assigned to the vertex finding efficiency: 
  $15\%$ for $y<0.06$, $7\%$ for $0.06<y<0.1$, $4\%$ for $0.1<y<0.2$
  and $1\%$ for $y>0.2$. 
  This efficiency is estimated using pseudo-CC data
  yielding an uncertainty of $2.4\%$ on the cross section measurements.

\item An uncertainty of $0.5\%$ accounts for the
  dependence of the acceptance correction on the PDFs used in the 
  MC simulation.

\item A $1.8\%$ uncertainty on the trigger efficiency is determined
  based on the pseudo-CC data sample. The uncorrelated component of
  this uncertainty is $1\%$.

\item An error of $0.8\%$ is estimated for
  the QED radiative corrections. This accounts for missing
  contributions in the simulation of the lowest order QED effects and
  for the uncertainty on the higher order QED and EW corrections.

\item In addition, there is a global uncertainty of $1.3\%$
 on the luminosity measurement for both the $R$ and $L$ data samples,
 of which $0.5\%$ is considered as correlated.

\end{itemize}
The total systematic error is formed by adding the individual
uncertainties in quadrature and amounts to about $4\%$ on
the cross section measurements.

The polarisation measurements have a relative uncertainty of $3.5\%$
for the TPOL~\cite{Lorenzon} and $1.6\%$ for the LPOL~\cite{lpol} polarimeter 
and yield an absolute uncertainty on the mean polarisation of $\pm0.7\,\%$
for the $R$ sample and $\pm1.1\,\%$ for the $L$ sample.  These are not
included in the total systematic error on the cross section
measurements, but are considered as independent uncertainties 
in a linear fit to the data.

\section{Results}

\begin{figure}[htb]
\begin{center}
\begin{picture}(90,150)(0,0)
\setlength{\unitlength}{1 mm}
\put(-30,  0){\epsfig{file=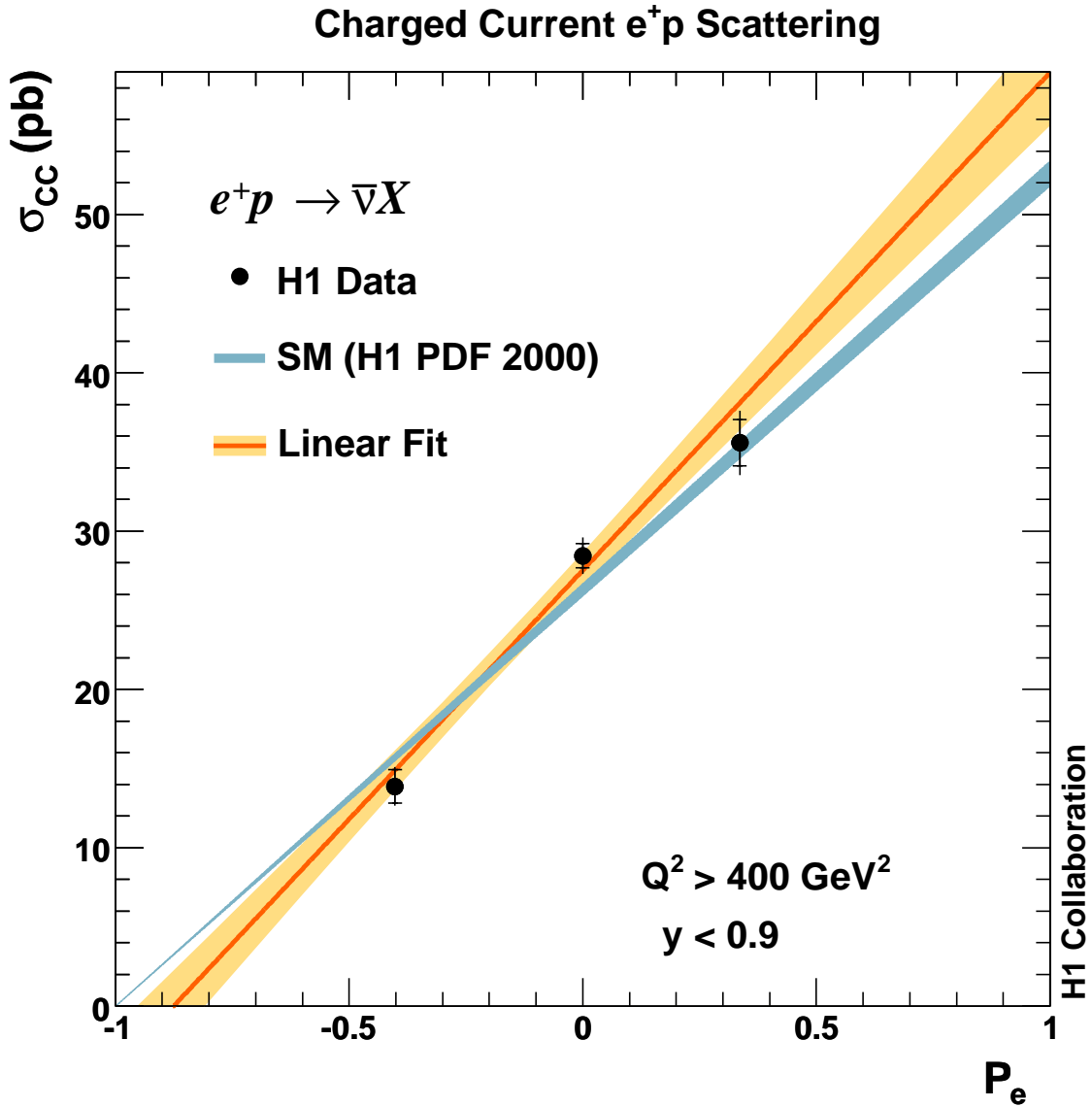,width=\textwidth}}
\end{picture}
\end{center}
\caption{
  The dependence of the $e^+p$ CC cross section on the lepton beam
  polarisation $P_e$. The inner and outer error bars
  represent respectively the statistical and total errors. The
  uncertainties on the polarisation measurement are smaller than the
  symbol size.  The data are compared to the Standard Model prediction
  based on the H1 PDF $2000$ parametrisation (dark shaded band). 
  The light shaded
  band corresponds to the resulting one-sigma contour of a linear fit
  to the data shown as the central line.}
\label{xsec}
\end{figure}

The measured integrated CC cross sections are quoted in the range
$Q^2> 400\,{\rm GeV}^2$ and $y<0.9$ and are given in Table~\ref{cctot}
and shown in Fig.~\ref{xsec}.  The measurement of the unpolarised
total cross section in the same phase space based on $65.2\,{\rm
pb}^{-1}$ of data collected in $1999$ and $2000$ is also given. This
measurement follows identically the procedure described
in~\cite{h1hiq2} but with the $Q^2$ cut adopted in this analysis. The
systematic uncertainties of this unpolarised measurement are taken to
be the same as in~\cite{h1hiq2}, with the exception of the QED
radiative correction uncertainty, which has been reduced from $3\%$ to
$0.8\%$. The measurements are compared to expectations of the Standard
Model using the H1 PDF $2000$ parametrisation. The uncertainty on the
Standard Model expectations combines the uncertainties from
experimental data used in the H1 PDF $2000$ fit as well as model
uncertainties~\cite{h1hiq2}.

\begin{table}[h]
  \begin{center}
    \begin{tabular}{|r|c|c|}
\hline
$P_e \,\,(\%)$ & $\sigma^{\rm tot}_{\rm CC} \,\,({\rm pb})$ &  SM expectation $({\rm pb})$\\
\hline
$+33.6$ & $35.6 \pm 1.5_{\rm stat} \pm 1.4_{\rm sys}$ & $35.1 \pm 0.6$ \\
$  0.0$ & $28.4 \pm 0.8_{\rm stat} \pm 0.8_{\rm sys}$ & $26.3 \pm 0.4$ \\
$-40.2$ & $13.9 \pm 1.1_{\rm stat} \pm 0.6_{\rm sys}$ & $15.7 \pm 0.3$ \\
\hline
\end{tabular} 
\caption{ \sl Measured cross section values for
    $\sigma^{\rm tot}_{\rm CC}(e^+p\rightarrow\overline{\nu}X)$ 
    in the region $Q^2> 400\,{\rm GeV}^2$ and 
    $y<0.9$ compared to the Standard Model (SM) expectation.}
\label{cctot}
\end{center}
\end{table}

A linear fit to the polarisation dependence of the measured cross
sections is performed taking into account the correlated systematic
uncertainties between the measurements and is shown in Fig.~\ref{xsec}.
The fit provides a reasonable description of the data with a
$\chi^2=4.4$ for one degree of freedom (dof). The result of the fit 
extrapolated to the point $P_e=-1$ yields a fully left handed charged current
cross section of
%
\begin{equation}
\label{result}
\sigma^{\rm tot}_{\rm CC} (P_e=-1)= -3.9 \pm 2.3_{\rm stat}  \pm 0.7_{\rm sys} 
\pm 0.8_{\rm pol}\,\, {\rm pb}\,,
\end{equation}
where the quoted errors correspond to the statistical (stat),
experimental (sys) and polarisation-related (pol) systematic
uncertainties.  This extrapolated cross section is consistent with the
Standard Model prediction of a vanishing cross section and corresponds
to an upper limit on $\sigma^{\rm tot}_{\rm CC} (P_e=-1)$ of $1.9\,{\rm pb}$ 
at $95\%$ confidence level (CL), as derived according to~\cite{feldman}. 
This result excludes the existence of charged currents involving
right handed fermions mediated by a boson of mass below $208\,{\rm GeV}$ at $95\%$\,CL,
assuming Standard Model couplings and a massless right handed $\nu_e$.

It is also possible to fit the measured cross sections by constraining the
cross section at $P_e=-1$ to zero. This yields a cross section at
$P_e=0$ of $27.5\pm 0.6_{\rm stat} \pm 0.9_{\rm sys}\,{\rm pb}$ with a
$\chi^2/{\rm dof}=3.5$ and a negligible error due to the uncertainty
on the polarisation measurement. The fitted value agrees well with the
Standard Model expectation of $26.3\pm 0.4\,{\rm pb}$.

\section{Summary}

The first measurement has been performed of polarised $e^+p$ total
charged current cross sections in the kinematic region of
$Q^2>400\,{\rm GeV}^2$ and $y<0.9$.  The results presented here are
based on data collected from collisions of unpolarised protons with
unpolarised positrons and, for the first time, with longitudinal
polarised positrons in left and right helicity states. The
polarisation dependence of the charged current cross section has thus been
established at HERA, extending previous tests of the chiral structure of
the charged current interaction into the region of large, space-like $Q^2$. 
The data are found to be consistent with the absence of right handed charged
currents as predicted by the Standard Model.

\section*{Acknowledgements}

We are grateful to the HERA machine group whose outstanding
efforts have made this experiment possible.
We thank the engineers and technicians for their work in constructing 
and maintaining the H1 detector, our funding agencies for
financial support, the DESY technical staff for continual assistance
and the DESY directorate for support and for the
hospitality which they extend to the non DESY
members of the collaboration.

\newpage

\end{document}